\documentclass[aps,prd,amssymb,nofootinbib,twocolumn,epsf,floatfix,
               superscriptaddress]{revtex4}
\usepackage[usenames]{color}
\usepackage[normalem]{ulem} 
\usepackage{graphicx} 
\usepackage{subfigure} 
\usepackage{amssymb}  
\usepackage{amsmath}
\usepackage{mathrsfs}
\usepackage{epstopdf}

\makeatletter
\renewcommand{\p@subfigure}{\thefigure}
\makeatother

\begin{document}
 

\title{Chebyshev Based Spectral Representations of Neutron-Star Equations
  of State}

\author{Lee Lindblom}
\affiliation{Department of Physics, University of California at San
  Diego}
\author{Tianji Zhou}
\affiliation{Department of Physics and Astronomy, Haverford College}

\date{\today}
 
\begin{abstract}
  Causal parametric representations of neutron-star equations of state
  are constructed here using Chebyshev polynomial based spectral
  expansions.  The accuracies of these representations are evaluated
  for a collection of model equations of state from a variety of
  nuclear-theory models and also a collection of equations of state
  with first- or second-order phase transitions of various sizes.
  These tests show that the Chebyshev based representations are
  convergent (even for equations of state with phase transitions) as
  the number of spectral basis functions is increased.  This study
  finds that the Chebyshev based representations are generally more
  accurate than a previously studied power-law based spectral
  representation, and that pressure-based representations are
  generally more accurate than those based on enthalpy.
\end{abstract} 
 
\maketitle

\section{Introduction}
\label{s:Introduction}

Parametric representations of the neutron-star equation of state are
used to model the poorly understood high-density material in the cores
of these stars.  The physical values of the parameters in these
equation of state models can be determined by matching the macroscopic
properties of the neutron-star models constructed from them
(e.g. their masses, radii, or tidal deformabilities) with astronomical
observations of those properties~\cite{Lindblom2014a}.  Fixing the
equation of state parameters in this way provides a determination of
the otherwise unobservable high-density neutron-star equation of
state.

Faithful representations of the equation of state must satisfy basic
thermodynamic stability and causality conditions along with minimal
accuracy requirements.  Thermodynamic stability requires the energy
density of the material to increase monotonically as the pressure is
increased.  Causality requires the sound speed determined by the
equation of state to be less than or equal to the speed of light.
Equation of state representations must also be accurate enough to
model any physical equation of state at a level commensurate with the
accuracy of the available astrophysical observations.  As the
observations of neutron stars improve over time, useful
representations should include a systematic way to match those
improvements, e.g. by increasing the number of adjustable parameters.
Suitable parametric representations should therefore be convergent in
the sense that their accuracies increase as the as the number of
parameters is increased.

A number of parametric representations of the neutron-star equation of
state have been introduced in recent years~\cite{Read:2008iy,
  Lindblom2010, Lindblom2018, Lindblom2022}.  All these
representations appear to be sufficiently accurate to accommodate the
precision of the currently available observations.  Some of these
representations ensure that the causality condition is satisfied.  And
some have been shown to be convergent in the sense that their
accuracies can be increased by increasing the number of parameters.
The most efficient representations, i.e. those providing the best
accuracy for a given number of parameters, are based on spectral
expansions~\cite{Lindblom2022}.

The physical neutron-star equation of state may or may not have
discontinuities caused by a phase transition.  Constructing accurate
parametric representations of equations of state with discontinuities
is particularly challenging.  The causal spectral representations that
provide the most accurate representations (for a given number of
parameters) of nuclear-theory based equations of state have recently
been shown to be non-convergent when used to represent equations of
state with strong phase transitions~\cite{Lindblom2024}.  The purpose
of this paper is to determine whether Chebyshev polynomial based
spectral expansions provide more robust and more accurate
representations of neutron-star equations of state, including those
with phase transitions.

Section~\ref{s:ChebyshevBasedRepresentations} defines causal
parametric representations of the neutron-star equation of state based
on Chebyshev polynomial spectral expansions.  These new
representations include both pressure- and enthalpy-based versions of
the equation of state. Section~\ref{s:NumericalTests} describes the
results of a series of numerical tests that measure the accuracy of
these new representations, including comparisons with the previously
studied spectral representations.  The model equations of state used
in these tests include a collection of nuclear-theory based
neutron-star equations of state and a collection of model equations of
state that include first- or second-order phase transitions.  The
implications of these results are discussed in
Sec.~\ref{s:Discussion}.

\section{Causal Chebyshev-Based Spectral Representations}
\label{s:ChebyshevBasedRepresentations}

Shortly after their formation, the temperatures in the cores of
neutron stars fall well below the local Fermi temperature, so thermal
contributions to the pressure and energy density become
negligible~\cite{Potekhin2015}.  The thermodynamic state of this
high-density material should therefore be well approximated by a
barotropic equation of state: $\epsilon=\epsilon(p)$, where $\epsilon$
is the total mass-energy density and $p$ is the pressure.

The speed of sound, $v$, in a barotropic fluid is determined by the
equation of state: $v^2=dp/d\epsilon$~\cite{Landau1959}.  These sound
speeds are causal if and only if the velocity function $\Upsilon$,
\begin{equation}
  \Upsilon=\frac{c^2-v^2}{v^2},
    \label{e:Upsilon_Def}
\end{equation}
is non-negative, $\Upsilon\geq 0$, where $c$ is the speed of light.

\subsection{Pressure-Based Spectral Expansions}
\label{s:PressureBasedExpansions}

The velocity function $\Upsilon$ is determined by the equation of
state: $\Upsilon(p)=c^2\,d\epsilon/dp -1$.  Conversely, $\Upsilon(p)$
can be used as a generating function from which the standard equation
of state, $\epsilon=\epsilon(p)$, can be determined by quadrature.
The procedure for determining $\epsilon=\epsilon(p)$ from
$\Upsilon(p)$ is summarized in
Appendix~\ref{s:CausalPressureBasedRepresentations}.

Causal parametric representations of the neutron-star equation of
state can be constructed by expressing $\Upsilon(p,\upsilon_a)$ as a
spectral expansion:
\begin{equation}
  \Upsilon(p,\upsilon_a)=\exp\left\{\sum_{a=0}^{N_\mathrm{parms}-1}
  \upsilon_a\Phi_a(p)\right\},
  \label{e:Upsilon_p_spec}
\end{equation}
where $\Phi_a(p)$ are the spectral basis functions and $\upsilon_a$
the spectral parameters.  These expansions guarantee that
$\Upsilon(p)\geq 0$ for every choice of $\upsilon_a$. Therefore any
equation of state determined from one of these
$\Upsilon(p,\upsilon_a)$ automatically satisfies the causality and
thermodynamic stability conditions.

This study explores the use of Chebyshev polynomial basis functions in
these spectral expansions:
\begin{equation}
  \Upsilon(p,\upsilon_a)=\Upsilon_0\exp\left\{
  \sum_{a=0}^{N_\mathrm{parms}-1}\upsilon_a(1+y)T_a(y)\right\},
  \label{e:Upsilon_p_Chebyshev}
\end{equation}
where the $T_a(y)$ are Chebyshev polynomials.  The variable $y$
(defined below) is a function of the pressure having the property that
$y=-1$ when $p=p_0$. The constants $p_0$ and $\Upsilon_0$ are
evaluated from the low-density equation of state at the point $p=p_0$
where it matches onto the high density spectral representation
determined by Eq.~(\ref{e:Upsilon_p_Chebyshev}).  Choosing $p_0$ and
$\Upsilon_0$ in this way ensures that no artificial first- or
second-order phase-transition discontinuity is introduced at the
matching point.

Chebyshev polynomials are defined by the recursion relation
$T_{a+1}(y)=2yT_a(y)-T_{a-1}(y)$ with $T_0(y)=1$ and $T_1(y)=y$.
Spectral expansions using Chebyshev basis functions are well behaved
on the domain $-1\leq y \leq 1$~\cite{Boyd1999}.  Therefore the
variable $y$ that appears in Eq.~(\ref{e:Upsilon_p_Chebyshev}) has
been defined as
\begin{equation}
  y = -1 + 2\log\left(\frac{p}{p_0}\right)\left[
    \log\left(\frac{p_\mathrm{max}}{p_0}\right)\right]^{-1},
  \label{e:zDef}
\end{equation}
to ensure that $-1\leq y\leq 1$ for pressures in the range $p_0\leq p
\leq p_\mathrm{max}$.  The factor $1+y$ that appears in
Eq.~(\ref{e:Upsilon_p_Chebyshev}) ensures that
$\Upsilon(p,\upsilon_a)$ has the limit,
$\Upsilon(p_0,\upsilon_a)=\Upsilon_0$, for every choice of spectral
parameters $\upsilon_a$.

\subsection{Enthalpy-Based Spectral Expansions}
\label{s:EnthalpyBasedExpansions}

For some purposes it is more convenient to use enthalpy-based
representations of the neutron-star equation of state.\footnote{The
standard Oppenheimer-Volkoff~\cite{Oppenheimer1939} representation of
the relativistic stellar structure equations has the property that
$dp/dr\rightarrow 0$ at the surface of the star.  This fact makes it
difficult to accurately determine the location of the star's surface
numerically.  The enthalpy based representation of these
equations~\cite{Lindblom1992} have the property that $dh/dr\rightarrow
-M/[R(R-2M)]$, making it easier to compute the star's radius
accurately in this case.} The enthalpy, $h(p)$, defined by
\begin{equation}
  h(p) = \int_0^p \frac{dp'}{\epsilon(p')\,c^2+p'},
  \label{e:enthalpyDef}
\end{equation}
is a monotonically increasing function of the pressure $p$.  Therefore
the velocity function $\Upsilon(p)$ defined in
Eq.~(\ref{e:Upsilon_Def}) can also be expressed as a function of the
enthalpy, $\Upsilon=\Upsilon(h)$.

Causal representations of the equation of state can also be generated
using enthalpy-based spectral expansions of the velocity function
$\Upsilon(h)$:
\begin{equation}
  \Upsilon(h,\upsilon_a)=\exp\left\{\sum_{a=1}^{N_\mathrm{parms}}\upsilon_a
  \Phi_a(h)\right\},
  \label{e:SpectralExpansion}
\end{equation}
where $\Phi_a(h)$ are a suitable set of enthalpy-based basis
functions. Any equation of state constructed in this way automatically
satisfies the causality and thermodynamic stability conditions:
$\Upsilon(h,\upsilon_a)\geq 0$.  The procedure for generating the
enthalpy-based equation of state, $\epsilon=\epsilon(h,\upsilon_a)$
and $p=p(h,\upsilon_a)$, from $\Upsilon(h,\upsilon_a)$ is summarized
in Appendix~\ref{s:CausalEnthalpyBasedRepresentations}.

This study explores the use of Chebyshev polynomials as spectral basis
functions:
\begin{equation}
  \Upsilon(h,\upsilon_a)=\Upsilon_0\exp\left\{\,
  \sum_{a=0}^{N_\mathrm{parms}-1}\,\upsilon_a\,\left(1+z\right)\,T_a(z)\right\},
  \label{e:Upsilon_spectral}
\end{equation}
where the $T_a(z)$ are Chebyshev polynomials and the variable $z$ is
given by
\begin{equation}
  z=-1+2\log\left(\frac{h}{h_0}\right)
  \left[\log\left(\frac{h_\mathrm{max}}{h_0}\right)\right]^{-1}.
  \label{e:yDef}
\end{equation}
The factor $1+z$ is included in Eq.~(\ref{e:Upsilon_spectral}) to
ensure that $\Upsilon(h_0,\upsilon_k)=\Upsilon_0$ for every choice of
$\upsilon_a$.

This study compares the accuracies of the Chebyshev polynomial based
spectral representations defined in Eq.~(\ref{e:Upsilon_spectral})
with those defined with the simple power-law spectral basis functions,
\begin{equation} 
  \Upsilon(h,\upsilon_a)=\Upsilon_0\exp\left\{\sum_{a=1}^{N_\mathrm{parms}}
    \upsilon_a\left[\log\left(\frac{h}{h_0}\right)\right]^a\right\},
    \label{e:PowerLawBasisDef}
\end{equation}
used in previous studies~\cite{Lindblom2018, Lindblom2022}.  This
study also compares the accuracy of the pressure-based Chebyshev
representations defined in Eq.~(\ref{e:Upsilon_p_Chebyshev}) with
the enthalpy-based representations defined in Eq.~(\ref{e:Upsilon_spectral}).

\section{Numerical Tests}
\label{s:NumericalTests}

This section describes a series of tests that measure the accuracy of
the Chebyshev based spectral representations described in
Sec.~\ref{s:ChebyshevBasedRepresentations}.  Best-fit spectral
representations are constructed and their accuracies evaluated
numerically for a variety of model neutron-star equations of state.
Three different collections of reference equations of state are used
in these tests.  The first reference collection consists of 26
nuclear-theory based neutron-star equations of state\footnote{The 26
nuclear-theory based equations of state used here are a causal subset
of those used by Read et al.~\cite{Read:2008iy} in their study of the
piecewise-polytropic representations of neutron-star equations of
state. The abbreviated names of these equations of state are: PAL6,
SLy, APR1, WFF3, BBB2, BPAL12, MPA1, MS1, MS1b, PS, GS1, GS2, BGN1H1,
GNH3, H1, H2, H3, H4, H5, H6, H7, PCL2, ALF2, ALF3. ALF4 and GM1L.
See Ref.~\cite{Read:2008iy} for descriptions of these nuclear-theory
models and the citations to the literature that define them.} used by
a number of studies to evaluate the accuracy of various parametric
representations~\cite{Read:2008iy, Lindblom2010, Lindblom2018,
  Lindblom2022}.  The second reference collection consists of
equations of state with discontinuities representing first-order phase
transitions having a range of sizes~\cite{Lindblom2024}.  These
equations of state were constructed by inserting a discontinuity into
the GM1L nuclear-theory based equation of state.\footnote{ The GM1L
equation of state was constructed in Ref.~\cite{spinella2017:PHD} from
the GM1 equation of state~\cite{Glendenning1991} by adjusting the
slope of the symmetry energy to agree with the established value,
$L=55$ MeV, using the formalism developed in Ref.~\cite{Typel2010}.}
The third reference collection is analogous to the second with
discontinuities representing second-order phase transitions inserted
with a range of sizes~\cite{Lindblom2024}.  The reference equations of
state used in these collections are represented as tables of enthalpy,
pressure, and energy density values: $\{h_i,p_i,\epsilon_i\}$ for
$1\leq i \leq N_\mathrm{table}$.

The numerical tests performed here construct spectral fits on the
domain $p_0 = 1.20788\times 10^{32}$ erg/cm${}^3 \leq p $ . The upper
limit of this domain, $p\leq p_\mathrm{max}$, is taken to be the
largest entry in the reference equation of state table.  The constant
$\Upsilon_0$ is determined by differentiating the interpolation
formula for the reference equation of state table.

\subsection{Enthalpy Based Tests}
\label{s:EnthalpyBasedTests}

A primary motivation for this study was the finding in
Ref.~\cite{Lindblom2024} that the enthalpy-based spectral
representations using power-law basis functions were not convergent
for equations of state with discontinuities caused by phase
transition.  Comparing the performance of the Chebyshev based
representations introduced in
Sec.~\ref{s:ChebyshevBasedRepresentations} with the power-law based
representations is therefore an important goal of this study.

Causal enthalpy-based spectral equation of state representations were
computed following the methods described in
Sec.~\ref{s:ChebyshevBasedRepresentations} and
Appendix~\ref{s:CausalEnthalpyBasedRepresentations}.  The integrals
used to construct the equation of state in
Appendix~\ref{s:CausalEnthalpyBasedRepresentations} were performed
numerically using Gaussian quadrature~\cite{numrec_f}.  The resulting
model equations of state are used to evaluate the energy densities,
$\epsilon(h_i,\upsilon_a)$, at the tabulated enthalpy values, $h_i$,
of the reference equations of state.  These model energy-density
values are then compared to the tabulated reference energy densities,
$\epsilon_i$, from the reference equation of state tables using the
error measure,
\begin{equation}
  \chi^2(\upsilon_a) = \frac{1}{N_\mathrm{table}}
  \sum_{i=1}^{N_\mathrm{table}}\left[\log\left(
    \frac{\epsilon(h_i,\upsilon_a)}{\epsilon_i}\right)\right]^2.
    \label{e:chi_enthalpy}
\end{equation}
The optimal or ``best-fit'' parametric representations are found by
minimizing $\chi^2(\upsilon_a)$ with respect to the spectral
parameters $\upsilon_a$.

The numerical calculations used in this study were performed using two
independent codes to confirm the accuracy of the results. The
minimizations of $\chi^2$ were carried out numerically using a Fortran
implementation of the Levenburg-Marquardt algorithm as described in
Ref.~\cite{numrec_f}, and using the scipy.optimize.least\_squares
implementation in Python~\cite{2020SciPy-NMeth}.  The resulting
minimum values of $\chi$ measure the accuracy of the best-fit
parametric representations.  These minimum values computed by the two
codes agree to within a few percent for the new Chebyshev based
representations.

\begin{figure}[!h]
    \includegraphics[width=0.4\textwidth]
                    {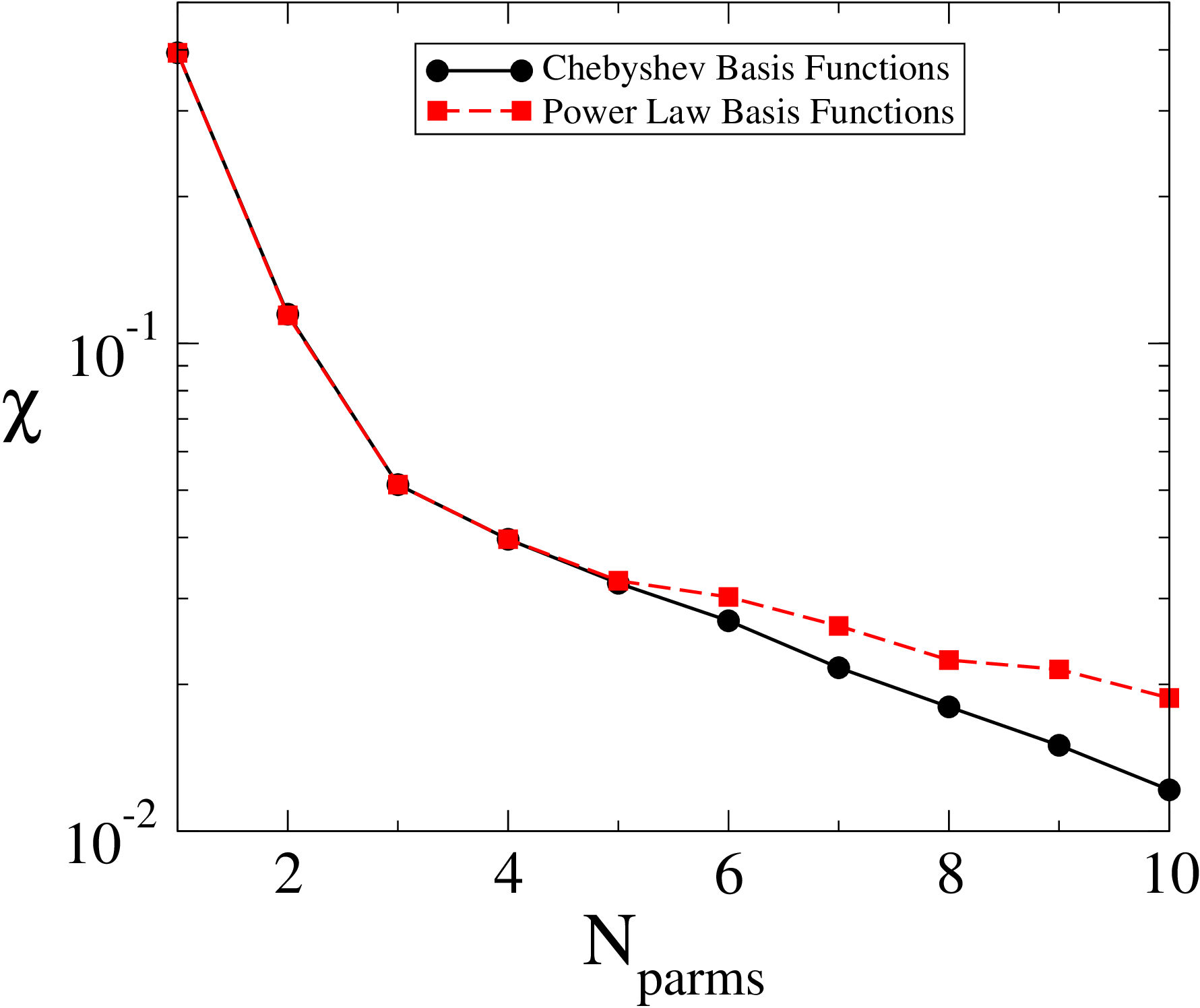}
      \caption{\label{f:CausalEOSAverages_h}Average modeling errors
        $\chi$ are illustrated as a function of $N_\mathrm{parms}$,
        the number of spectral parameters, for the reference
        collection of 26 nuclear-theory based neutron-star equation of
        state models.  The solid (black) curve gives results for the
        enthalpy-based Chebyshev basis functions, while the dashed
        (red) curve gives results for the simpler enthalpy-based
        power-law basis functions. }
\end{figure}

Figure~\ref{f:CausalEOSAverages_h} illustrates the errors, $\chi$, as
a function of the number of spectral parameters, $N_\mathrm{parms}$,
for the best-fit models averaged over the reference collection of 26
nuclear-theory based neutron star equations of state. The Chebyshev
and power-law based representations have almost identical errors for
$N_\mathrm{parms}\leq 5$.  But the Chebyshev based models are more
accurate than the power-law based models for larger
$N_\mathrm{parms}$.  The Chebyshev based models show faster and
cleaner exponential convergence than the power-law bases models for
these nuclear-theory based reference equations of state.

\begin{figure}[!h]
    \includegraphics[width=0.4\textwidth]
                    {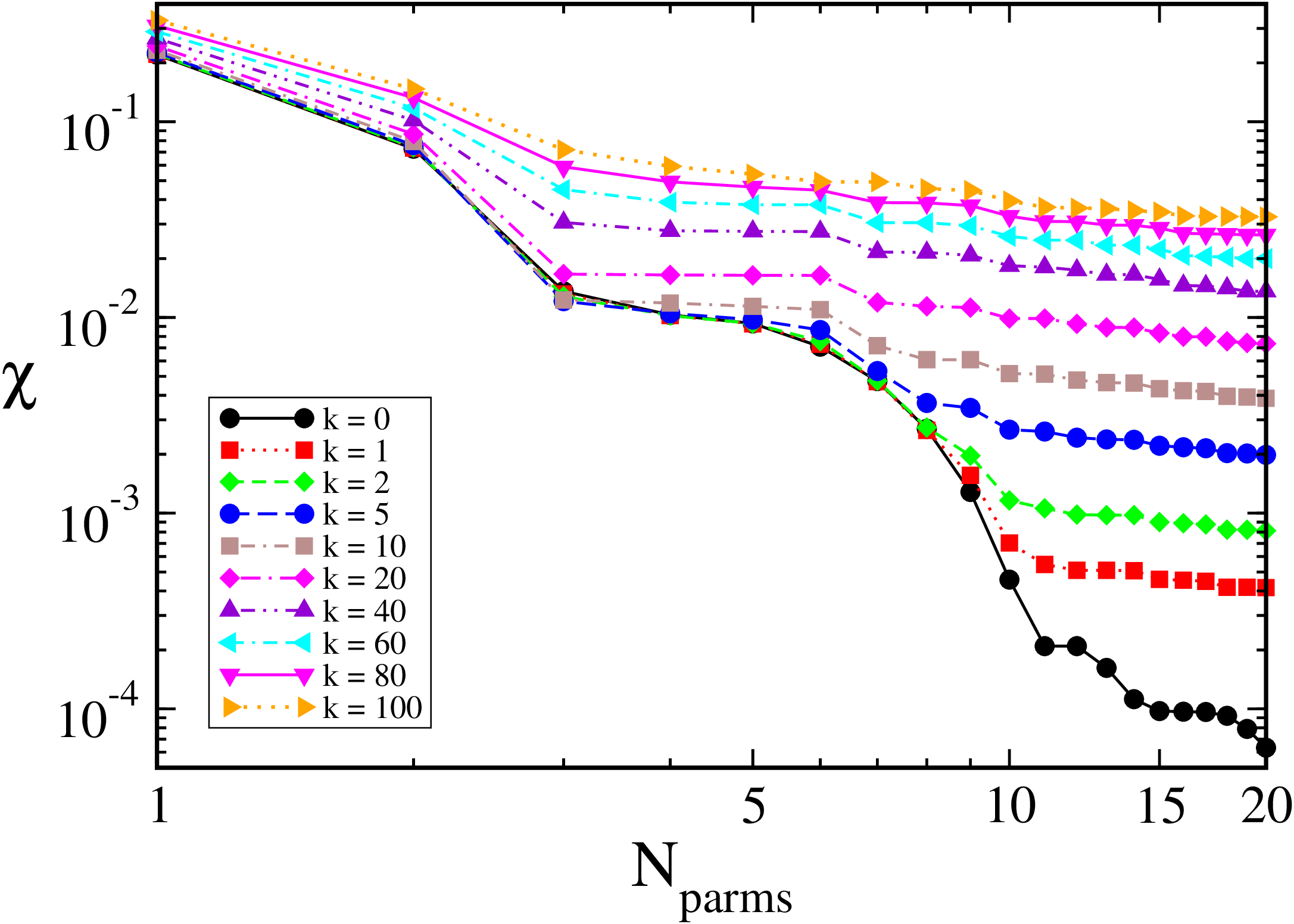}
      \caption{\label{f:FOChebyshevChi_h}Modeling errors $\chi$ for
        best-fit enthalpy-based Chebyshev spectral representations are
        illustrated as a function of $N_\mathrm{parms}$ for a
        collection of equations of state with first-order phase
        transitions.  The $k$ parameter is proportional to the size of
        the discontinuity caused by the phase transition, with $k=0$
        representing no discontinuity and $k=100$ the maximum
        discontinuity allowed for stable neutron stars.}
\end{figure}

\begin{figure}[!h]
    \includegraphics[width=0.4\textwidth]
                    {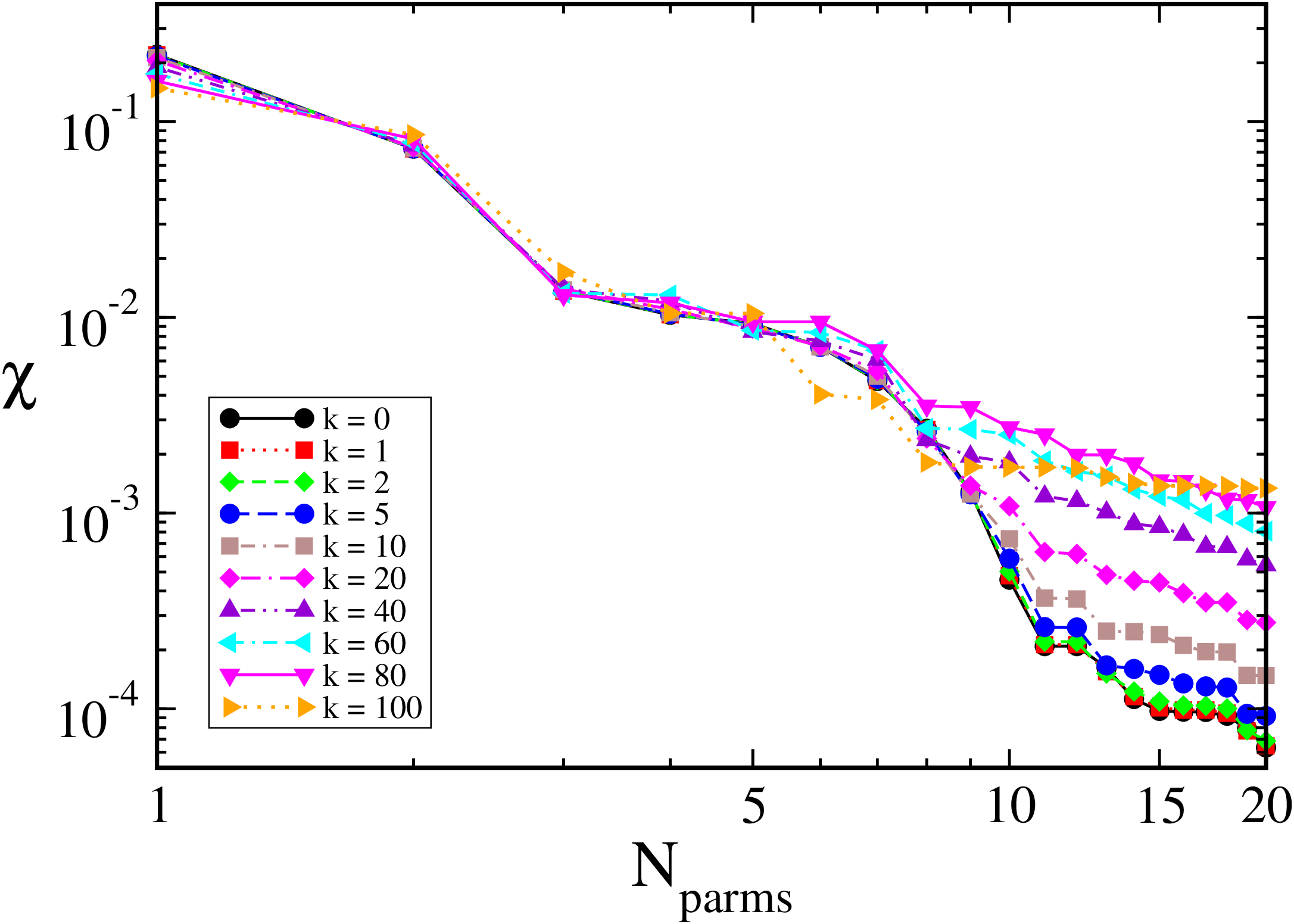}
      \caption{\label{f:SOChebyshevChi_h}Modeling errors $\chi$ for
        best-fit enthalpy-based Chebyshev spectral representations are
        illustrated as a function of $N_\mathrm{parms}$ for a
        collection of equations of state with second-order phase
        transitions of various sizes: $0\leq k \leq 100$. }
\end{figure}

Figures~\ref{f:FOChebyshevChi_h} and \ref{f:SOChebyshevChi_h}
illustrate the best-fit modeling errors $\chi$ as functions of
$N_\mathrm{parms}$ for the enthalpy-based Chebyshev polynomial
spectral representations of the reference equations of state with
first- or second-order phase transitions respectively.  The individual
curves in these figures represent equations of state with
phase-transition discontinuities having sizes proportional to the
parameter $k$.  The $k=0$ curves represent the GM1L equation of state
with no discontinuity added, while the $k=100$ curves correspond to
the equations of state with the maximum physically allowed
discontinuities, see Ref.~\cite{Lindblom2024} for details.

The linearity of the $\chi(N_\mathrm{parms})$ curves for large
$N_\mathrm{parms}$ in the log-log plots in
Figs.~\ref{f:FOChebyshevChi_h} and \ref{f:SOChebyshevChi_h} illustrate
the convergence of the best-fit enthalpy-based Chebyshev spectral
representations. The fitting errors for the equations of state with
first-order phase transitions shown in Fig.~\ref{f:FOChebyshevChi_h}
decrease with increasing $N_\mathrm{parms}$ as
$\chi(N_\mathrm{parms})\propto N_\mathrm{parms}^{-1/2}$
(approximately) for large $N_\mathrm{parms}$, while the fitting errors
for the equations of state with second-order phase transitions shown
in Fig.~\ref{f:SOChebyshevChi_h} decrease as
$\chi(N_\mathrm{parms})\propto N_\mathrm{parms}^{-3/2}$
(approximately).  The faster convergence of $\chi(N_\mathrm{parms})$
by an additional power of $N_\mathrm{parms}^{-1}$ for the equations of
state with the smoother second-order phase transitions is consistent
with the expectations for algebraically convergent Chebyshev spectral
expansions~\cite{Boyd1999}.

Enthalpy-based spectral representations using the simple power-law
basis functions were studied in Ref.~\cite{Lindblom2024} for the
reference equations of state with phase transition discontinuities.
The graphs of $\chi(N_\mathrm{parms})$ from that study did not show
convergence for the equations of state with larger discontinuities.
The detailed convergence graphs, analogous to
Figs.~\ref{f:FOChebyshevChi_h} and \ref{f:SOChebyshevChi_h}, are given
in Figs.~5 and 6 of Ref.~\cite{Lindblom2024}, and will not be repeated
here.  The minimum $\chi$ values for the non-convergent power-law
based representations computed by the two codes used in this study
agree qualitatively. But the differences in the error measures for the
two codes are about an order of magnitude larger for the power-law
based compared those for the Chebyshev based representations.

\begin{figure}[!h]
    \includegraphics[width=0.4\textwidth]
                    {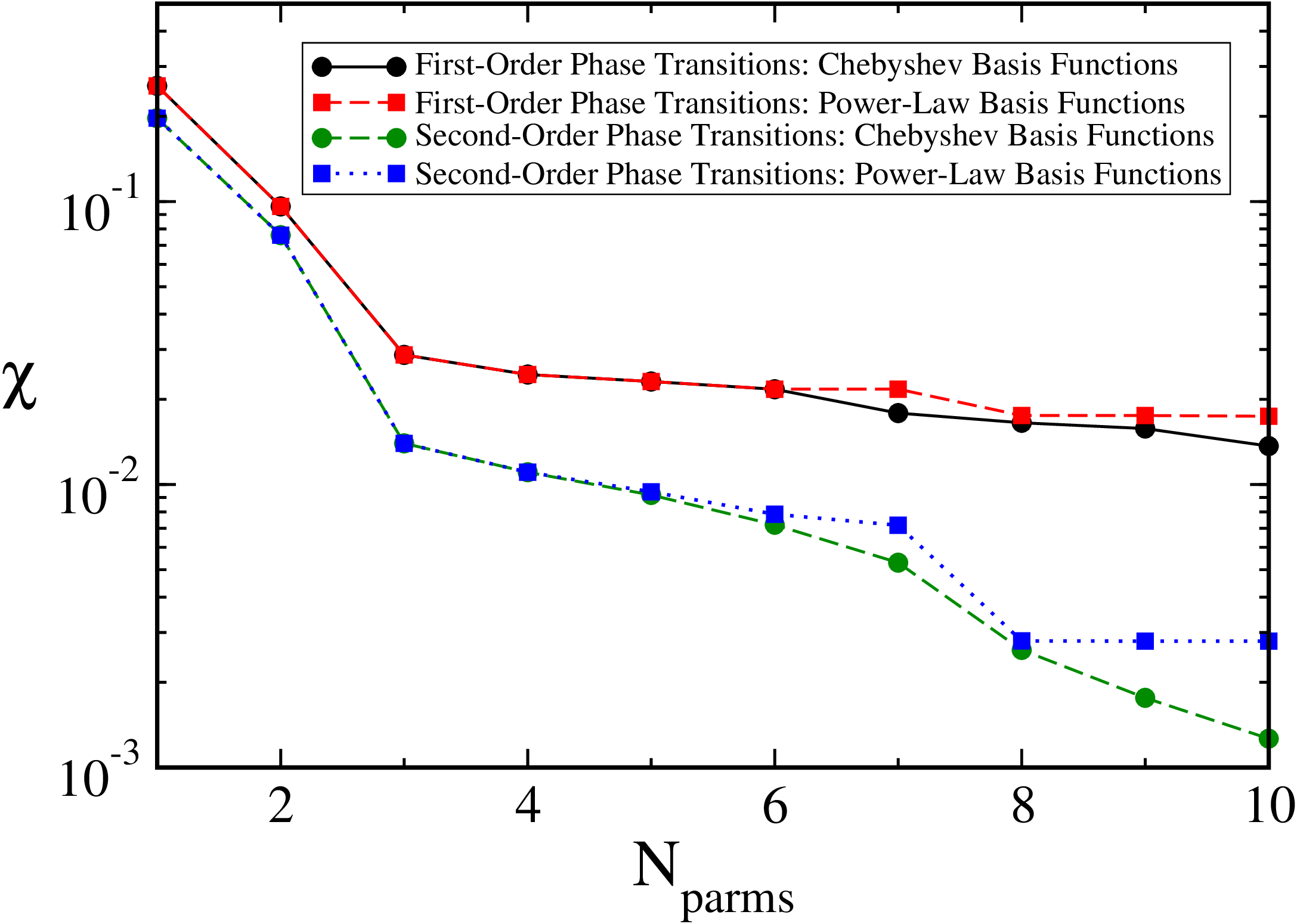}
      \caption{\label{f:EnthalpyBasedAverages}Average modeling errors
        $\chi$ using enthalpy-based spectral expansions are
        illustrated as a function of $N_\mathrm{parms}$, the number of
        spectral parameters, for a collection of equations of state
        with first- and second-order phase transitions of various
        sizes, $0\leq k\leq 100$.}
\end{figure}

Figure~\ref{f:EnthalpyBasedAverages} illustrates the relative
accuracies of the enthalpy-based Chebyshev and the power-law spectral
representations for averages over the reference collections of first-
and second-order phase transitions.  This figure illustrates the
non-convergence of the power-law representations for the larger values
of $N_\mathrm{parms}$.  And these results show that the Chebyshev
based spectral representations are somewhat more accurate than the
power-law representations for larger values of $N_\mathrm{parms}$
%

\subsection{Pressure Based Tests}
\label{s:PressureBasedTests}

Pressure-based spectral equations of state were computed following the
method described in Sec.~\ref{s:ChebyshevBasedRepresentations} and
Appendix~\ref{s:CausalPressureBasedRepresentations}.  The resulting
model equations of state are used to evaluate the energy densities,
$\epsilon(p_i,\upsilon_a)$, at the tabulated pressure values, $p_i$,
of the reference equations of state.  These model energy-density values
are then compared to the tabulated reference energy densities,
$\epsilon_i$, from the reference equation of state tables using the
error measure,
\begin{equation}
  \chi^2(\upsilon_a) = \frac{1}{N_\mathrm{table}}
  \sum_{i=1}^{N_\mathrm{table}}\left[\log\left(
      \frac{\epsilon(p_i,\upsilon_a)}{\epsilon_i}\right)\right]^2.
    \label{e:chi_pressure}
\end{equation}
The best-fit representations are determined by minimizing
$\chi(\upsilon_a)$ over the spectral parameters $\upsilon_a$.  The
resulting best-fit error measures $\chi(N_\mathrm{parms})$ for the
pressure-based Chebyshev representations are compared in this section
to the best-fit results for the enthalpy-based Chebyshev
representations.

\begin{figure}[!h]
    \includegraphics[width=0.4\textwidth]
                    {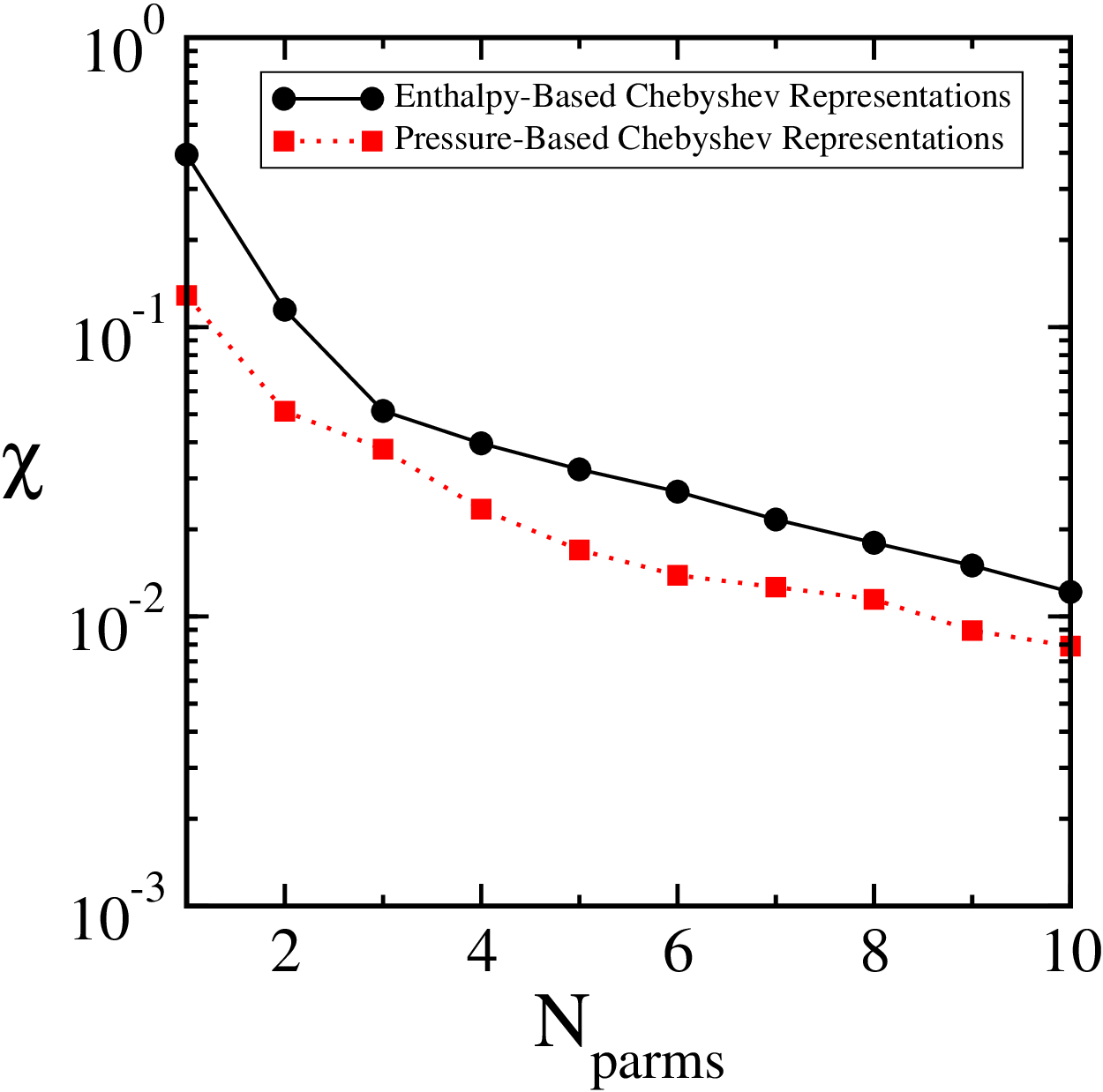}
      \caption{\label{f:PressureEnthalpyAveragesNucleartheory}Average
        modeling errors $\chi$ using pressure-based Chebyshev
        representations compared to those using enthalpy-based
        Chebyshev representations of 26 nuclear-theory based equation
        of state models.}
\end{figure}

Figure~\ref{f:PressureEnthalpyAveragesNucleartheory} illustrates the
average errors, $\chi$, as a function of the number of spectral
parameters, $N_\mathrm{parms}$, obtained for the best-fit models of
the reference collection of 26 nuclear-theory based neutron star
equations of state.  The (black) solid curve represents the results
using the enthalpy-based Chebyshev spectral representations, while the
(red) dotted curve represents the results using the pressure-based
Chebyshev spectral representations.  These results show that the
pressure-based representations have average errors roughly half those
of the enthalpy-based representations for each value of
$N_\mathrm{parms}$ in this collection of nuclear-theory based
equations of state.

\begin{figure}[!h]
    \includegraphics[width=0.4\textwidth]
                    {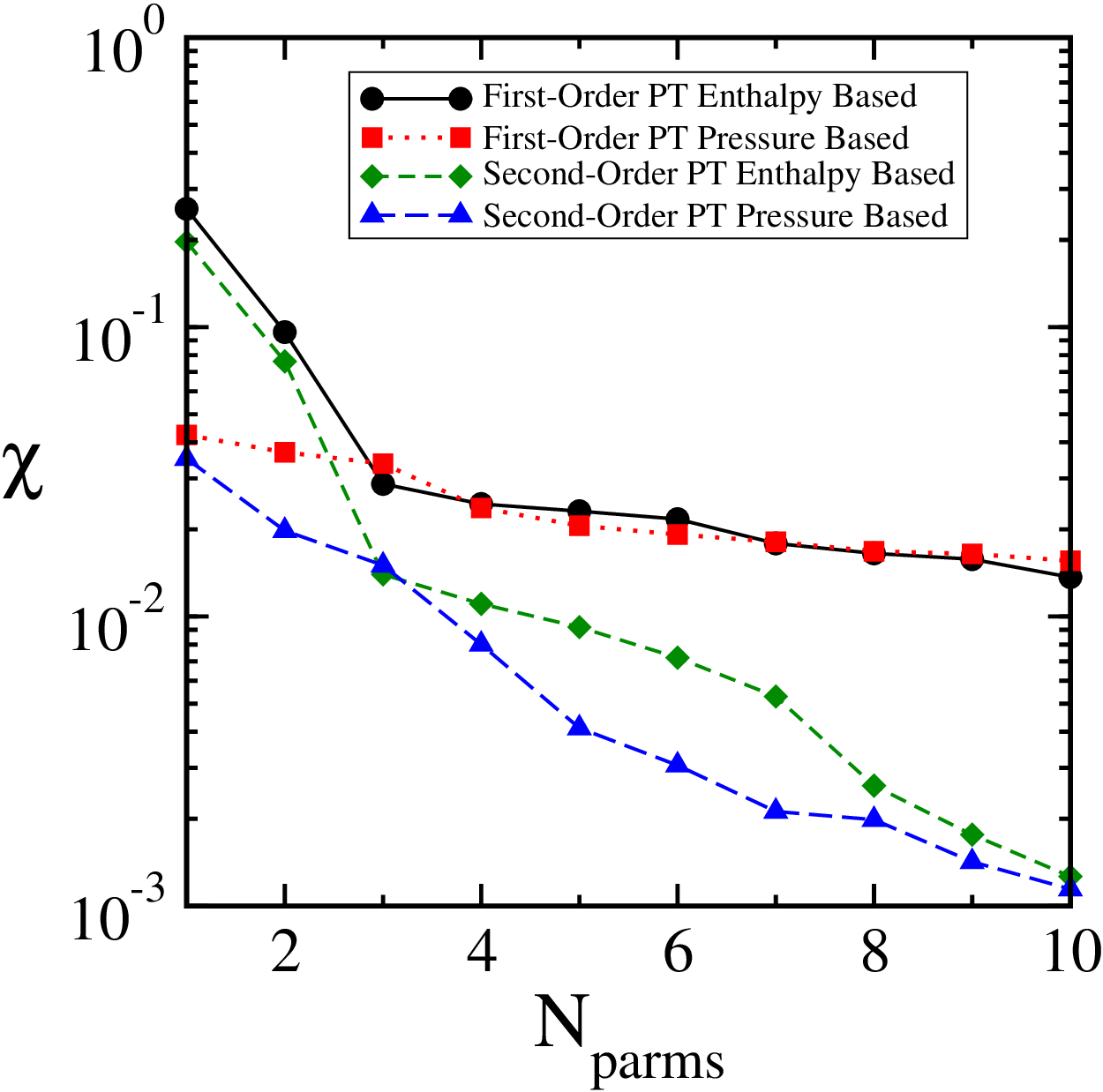}
      \caption{\label{f:PressureEnthalpyAverages}Average modeling
        errors $\chi$ using pressure-based Chebyshev spectral
        representations compared to those using enthalpy-based
        Chebyshev spectral representations.  Comparisons are given for
        the averages over the collections of reference equations of
        state with first- or second-order phase transitions of various
        sizes.}
\end{figure}

The pressure-based spectral expansions using Chebyshev basis functions
are convergent.  The detailed plots of $\chi(N_\mathrm{parms})$ for
the reference collections of equations of state with first- or
second-order phase transitions are qualitatively similar to
Figs.~\ref{f:FOChebyshevChi_h} and \ref{f:SOChebyshevChi_h} so they
will not be included here.  Figure~\ref{f:PressureEnthalpyAverages}
illustrates the average values of $\chi(N_\mathrm{parms})$ for the
collections of equations of state with first- or second-order phase
transitions.  As was the case for the enthalpy-based representations,
the fitting errors for the equations of state with first-order phase
transitions decrease with increasing $N_\mathrm{parms}$ as
$\chi(N_\mathrm{parms})\propto N_\mathrm{parms}^{-1/2}$
(approximately), while the fitting errors for the equations of state
with second-order phase transitions decrease as
$\chi(N_\mathrm{parms})\propto N_\mathrm{parms}^{-3/2}$
(approximately).  Figure~\ref{f:PressureEnthalpyAverages} shows that
the pressure-based Chebyshev representations are significantly better
than the enthalpy-based representations for $N_\mathrm{parms} < 3$.
For larger values of $N_\mathrm{parms}$ the pressure-based
representations have more or less the same accuracy as the
enthalpy-based representations of the equations of state with
first-order phase transitions.  The pressure-based representations are
uniformly better than the enthalpy-based representations with
$N_\mathrm{parms} >3$ for equations of state with second-order phase
transitions.
%

\section{Discussion}
\label{s:Discussion}

This study focused on two questions: Are the enthalpy-based spectral
expansions using Chebyshev polynomial basis functions an improvement
over the non-convergent previously studied expansions using power-law
basis functions?  How does the accuracy of the pressure-based
Chebyshev representations compare to the accuracy of the
enthalpy-based Chebyshev representations?

The tests summarized in Sec.~\ref{s:EnthalpyBasedTests} show that the
enthalpy-based spectral representations using Chebyshev polynomial
basis functions are generally more accurate those those using
power-law basis functions.  The Chebyshev based representations are
shown to be convergent where the power-law based representations were
not.  When applied to equations of state with discontinuities caused
by phase transitions, the convergence rates of the Chebyshev based
representations become algebraic at rates appropriate for spectral
representations.

The tests summarized in Sec.~\ref{s:PressureBasedTests} show that the
pressure-based spectral representations using Chebyshev polynomial
basis functions are (almost) uniformly better than the analogous
enthalpy-based representations.  These pressure-based representations
are convergent with the same convergence rates as the enthalpy-based
representations.  However the pressure-based representations have
average modeling errors that are roughly half those of the
enthalpy-based representations for the reference collection of 26
nuclear-theory based neutron-star equations of state.  The
pressure-based representations with $N_\mathrm{parms}<3$ are also
significantly more accurate than the enthalpy-based representations
for the equations of state with first- or second-order phase
transitions.

\appendix

\section{Causal Pressure-Based Representations}
\label{s:CausalPressureBasedRepresentations}

This appendix summarizes the procedure developed in
Ref.~\cite{Lindblom2018} for using the velocity function $\Upsilon(p)$
defined in Eq.~(\ref{e:Upsilon_Def}) as a generating function that
determines the pressure-based equation of state:
$\epsilon=\epsilon(p)$.

The definition of the velocity function $\Upsilon(p)$ can be
re-written as the ordinary differential equation,
\begin{equation}
  \frac{d\epsilon(p)}{dp} =\frac{1}{c^2}+\frac{\Upsilon(p)}{c^2}.
  \label{e:Upsilon_p_Def}
\end{equation}
This equation can be integrated to determine the pressure-based
equation of state, $\epsilon=\epsilon(p)$:
\begin{equation}
  \epsilon(p)=\epsilon_0+\frac{p-p_0}{c^2} 
  +\frac{1}{c^2}\int_{p_0}^p\Upsilon(p')dp'.
  \label{e:epsilon_p}
\end{equation}
Causal spectral representations of the equation of state can then be
constructed using Eq.~(\ref{e:epsilon_p}) with the pressure-based
representations of $\Upsilon(p,\upsilon_a)$ given in
Sec.~\ref{s:ChebyshevBasedRepresentations}.

\section{Causal Enthalpy-Based Representations}
\label{s:CausalEnthalpyBasedRepresentations}

This appendix summarizes the procedure developed in
Ref.~\cite{Lindblom2018} for using the velocity function $\Upsilon(h)$
defined in Eq.~(\ref{e:Upsilon_Def}) as a generating function that
determines the enthalpy-based equation of state:
$\epsilon=\epsilon(h)$ and $p=p(h)$.  Enthalpy based representations
of the equation of state are more useful than the standard
pressure-based representations for some purposes.  For example these
representations make it easier to solve the more accurate and more
efficient enthalpy-based representations of the Oppenheimer-Volkoff
neutron-star structure equations~\cite{Lindblom1992}.

The definition of the enthalpy, $h$, in Eq.~(\ref{e:enthalpyDef})
implies an expression for $dp/dh=\epsilon\,c^2 + p$.  Similarly, the
definition of the velocity function, Eq.~(\ref{e:Upsilon_Def}),
provides an expression for $d\epsilon/dh$:
\begin{equation}
  \Upsilon(h)=c^2 \frac{d\epsilon}{dp}-1 =
  c^2 \frac{d\epsilon}{dh}\,\left[\epsilon(h)\,c^2+p(h)\right]^{-1}-1.
  \label{e:Upsilon_h_Def}
\end{equation}
Together these definitions provide a system of first-order ordinary
differential equations for $\epsilon(h)$ and $p(h)$,
\begin{eqnarray}
  \frac{dp}{dh} &=& \epsilon\,c^2+p.
  \label{e:p_h_ode}\\
  \frac{d\epsilon}{dh}&=& \left(\epsilon+\frac{p}{c^2}\right)
  \left[\Upsilon(h)+1\right],
  \label{e:epsilon_h_ode}
\end{eqnarray}
These equations can be reduced to quadratures:
\begin{eqnarray}
  p(h) &=& p_0 + (\epsilon_0\,c^2+p_0)\int_{h_0}^h\mu(h')\,dh',
  \label{e:p_h_integral}\\
  \epsilon(h) &=& -\frac{p(h)}{c^2}
  + \left(\epsilon_0+\frac{p_0}{c^2}\right)\,\mu(h),
  \label{e:epsilon_h_integral}
\end{eqnarray}
where $p_0=p(h_0)$ and $\epsilon_0=\epsilon(h_0)$ represent a point on
the equation of state curve, and $\mu(h)$ is given by
\begin{equation}
  \mu(h)= \exp\left\{
  \int_{h_0}^h\left[2+\Upsilon(h')\right]\,dh'\right\}.
  \label{e:mu_h_integral}
\end{equation}
Equations~(\ref{e:p_h_integral})--(\ref{e:mu_h_integral}) determine a
causal spectral representation of the equation of state
for any of the enthalpy-based spectral expansions $\Upsilon(h,\upsilon_a)$
defined in Sec.~\ref{s:ChebyshevBasedRepresentations}.

\acknowledgments

We thank Fridolin Weber for providing us with the GM1L equation of
state table used in this study, including the literature citations to
its provenance.  This research was supported in part by the National
Science Foundation grants 2012857 and 2407545 to the University of
California at San Diego, USA. T.Z. was supported by a Haverford
College KINSC Summer Scholars fellowship.

\bibstyle{prd}
\bibliography{../References/References}

\end{document}